\documentclass[11pt]{article}
\title{No production of entropy in the Euler fluid}
\author{R. F. Streater, Dept. of Mathematics\\
King's College London, Strand,\\London WC2R 2LS}
\date{4 Oct 2003}

\newtheorem{theorem}{Theorem}
\begin{document}
\setlength{\oddsidemargin}{0in} \setlength{\evensidemargin}{0in}
\setlength{\topmargin}{-.4in} \maketitle
\begin{abstract}
We derive the Euler equations as the hydrodynamic limit of
a stochastic of a hard-sphere gas. We show that the system does 
not produce entropy. 

\end{abstract}
\section{Hydrostatics of a gas of hard spheres}
We take space to be $\Lambda\subseteq(a{\bf Z})^3$, and suppose
the length $a$, representing the diameter of a molecule, to be so
small compared with the variation of the macroscopic fields that
we can replace all sums over $\Lambda$ by integrals. The possible
configurations of the fluid are the points in the product sample
space
\[
\Omega=\prod_{\mbox{\boldmath$x$}\in\Lambda}\Omega_{\mbox{\boldmath$x$}},\]
so a configuration is specified by the collection
$\{\omega_{\mbox{\boldmath$x$}}\}_{\mbox{\boldmath$x$}\in\Lambda}$.
For each {\boldmath$x$},
\[
\Omega_{\mbox{\boldmath$x$}}=\left\{\emptyset,(\epsilon{\bf
Z})^3\right\}.\] Here, $\epsilon$ is a small parameter having the
dimension of momentum. If the system is in a configuration
$\omega$, such that $\omega_{\mbox{\boldmath$x$}}=\emptyset$, then
we say that the site {\boldmath$x$} is empty. If
$\omega_{\mbox{\boldmath$x$}}=\mbox{\boldmath$k$}$, we say that
the site {\boldmath$x$} is occupied, by a particle of momentum
{\boldmath$k$}. This simple exclusion of more than one particle on
each site incorporates the hard-core repulsion between the
particles, which are thus hard spheres sitting at some of the
points of $\Lambda$.

The {\em state} of the system is a probability on $\Omega$,
denoted by $\mu$. We denote the set of states by $\Sigma$. The
`slow variables' of our model are the 5 extensive conserved random
fields
\begin{eqnarray*}
{\cal N}_{\mbox{\boldmath$x$}}(\omega)&=&\left\{\begin{array}{ll}
               0&\mbox{ if }\omega_{\mbox{\boldmath$x$}}=\emptyset\\
               1&\mbox{ if }\omega_{\mbox{\boldmath$x$}}=\mbox{\boldmath$k$}
               \end{array}\right.\\
{\cal E}_{\mbox{\boldmath$x$}}(\omega)&=&\left\{\begin{array}{ll}
               0&\mbox{ if }\omega_{\mbox{\boldmath$x$}}=\emptyset\\
               \mbox{\boldmath$k\cdot k$}/2m+\Phi(\mbox{\boldmath$x$})
&\mbox{ if }\omega_{\mbox{\boldmath$x$}}=\mbox{\boldmath$k$}
               \end{array}\right.\\
\mbox{\boldmath${\cal P}_x$}(\omega)&=&\left\{\begin{array}{ll}
               0&\mbox{ if }\omega_{\mbox{\boldmath$x$}}=\emptyset\\
               \mbox{\boldmath$k$}&\mbox{ if
               }\omega_{\mbox{\boldmath$x$}}
=\mbox{\boldmath$k$}
               \end{array}\right.
\end{eqnarray*}
Here, $\Phi(\mbox{\boldmath$x$})$ is the external potential energy
per particle. The slow variables appearing in hydrodynamics are
the $n=5|\Lambda|$ means in the state $\mu$:
\[
N_{\mbox{\boldmath$x$}}={\bf E}_\mu[{\cal
N}_{\mbox{\boldmath$x$}}];
\hspace{.3in}E_{\mbox{\boldmath$x$}}={\bf E}_\mu[{\cal
E}_{\mbox{\boldmath$x$}}];\hspace{.3in}
\mbox{\boldmath$\Pi_x$}={\bf E}_\mu[\mbox {\boldmath${\cal
P}_x$}].\] The (von Neumann) entropy of any state $\mu$ is
\begin{equation}\label{entropy}
S(\mu):=-k_{_B}\sum_\omega\mu(\omega)\log\mu(\omega).
\end{equation}

In information geometry, the choice of slow variables
$\{X_1,X_2\ldots,X_n\}$ defines the {\em information manifold}
${\cal M}$, which consists of states of maximum entropy among all
states with given means, say
\[{\bf E}_\mu[X_j]:=
\mu\cdot X_j=\eta_j, \hspace{1in}j=1,\ldots n;\] By the
Gibbs-Jaynes principle, such a state has the form
\[\mu(\omega)=\exp\left\{-\sum_{j=1}^n \xi_j X_j(\omega)-\log\Xi\right\}\]
where the dual, or {\em canonical}, variables $\xi_j$ are Lagrange
multipliers determined uniquely by the means $\eta_i$. In our
case, for each {\boldmath$x$}, the duals to energy, number, and
momentum are, respectively,
$\beta_{\mbox{\boldmath$x$}},\xi_{\mbox{\boldmath$x$}},\mbox{\boldmath$\zeta
_x$}$, and so the state has the form
\begin{equation}
\mu(\omega)=\prod_{\mbox{\boldmath$x$}}\Xi_{\mbox{\boldmath$x$}}
^{-1}\exp\left\{-\xi_{\mbox{\boldmath$x$}}{\cal N}_
{\mbox{\boldmath$x$}}(\omega)-\beta_{\mbox{\boldmath$x$}}{\cal
E}_{\mbox{\boldmath$x$}}(\omega)-
\mbox{\boldmath$\zeta_x\cdot{\cal P}_x$}(\omega)\right\}.
\label{gce}
\end{equation}
Such a state is said to be in {\em local thermodynamic
equilibrium}, {\em LTE}. Equilibrium holds when $\beta...$ are
independent of {\boldmath$x$}. In \cite{RFS1} we found
the (grand) partition function at point {\boldmath$x$},
\[
\Xi_{\mbox{\boldmath$x$}}=1+\left(\frac{2\pi
m}{\epsilon^2\beta_{\mbox{\boldmath$x$}}}\right)^{3/2}\exp
\left\{-\xi_{\mbox{\boldmath$
x$}}-\beta_{\mbox{\boldmath$x$}}\Phi(\mbox{\boldmath$x$})+m\mbox{
\boldmath$\zeta_x\cdot\zeta_x$}/2\beta_{\mbox{\boldmath$x$}}\right\}
\] by replacing the sum over the momentum lattice of size
$\epsilon$ by a Gaussian integral. The product structure of an
{\em LTE} state means that an observable at a point of $\Lambda$
is independent of an observable at any other. Note that
$\mu_{\mbox{\boldmath$x$}}\{\emptyset\}=(1-N_{\mbox{\boldmath$x$}})$.
It then follows from (\ref{gce}) that
\begin{equation}
1-N_{\mbox{\boldmath$x$}}=\Xi^{-1}.
\end{equation}
If
$\omega_{\mbox{\boldmath$x$}}\neq \emptyset$, the state $\mu$ can
be written in Maxwell form
\[
\mu(\mbox{\boldmath$x,k$})=N_{\mbox{\boldmath$x$}}
p(\mbox{\boldmath$x,k$})=N_{\mbox{\boldmath$x$}}
Z^{-1}\exp\{-\beta_{\mbox{\boldmath$x$}}\Phi(\mbox{\boldmath$x$})-
\beta_{\mbox{\boldmath$x$}}\mbox{\boldmath$k\cdot
k$}/(2m)-\mbox{\boldmath$ \zeta_x\cdot k$}\}
\]
 where $Z$ is the canonical partition function:
\begin{equation}
Z_x=\left(\frac{2\pi
m}{\epsilon^2\beta_{\mbox{\boldmath$x$}}}\right)^{3/2} \exp\left\{-\beta_
{\mbox{\boldmath$x$}}\Phi(\mbox{\boldmath$x$})+ \frac{m\mbox
{\boldmath$\zeta_x\cdot\zeta_x$}}
{2\beta_{\mbox{\boldmath$x$}}}\right\}. \label{Z}
\end{equation}
We note the identity for each {\boldmath$x$}
\[\Xi=1+e^{-\xi}Z=1+e^{-\xi-\beta\Phi}Z_0,\]
where $Z_0$ is the canonical partition function when $\Phi=0$. The
external potential does not influence the local velocity
distribution $p$, as it is cancelled out by the partition
function.
 The mean fields are related to the canonical fields
by
\begin{eqnarray*}
E_{\mbox{\boldmath$x$}}=-\frac{\partial}{\partial
\beta_{\mbox{\boldmath$x$}}}
\log\Xi_{\mbox{\boldmath$x$}}&=&N(\mbox{\boldmath$x$})
\left(\Phi(\mbox{\boldmath$x$})+\frac{3}
{2\beta_{\mbox{\boldmath$x$}}}+\frac{m\mbox{\boldmath$\zeta_x\cdot\zeta_x$}}
{2\beta_{\mbox{\boldmath$x$}}^2}\right)\\
N_{\mbox{\boldmath$x$}}=-\frac{\partial}{\partial\xi_{\mbox{\boldmath$x$}}}
\log\Xi_{\mbox{\boldmath$x$}}&=&\frac{\Xi_{\mbox{\boldmath$x$}}-1}
{\Xi_{\mbox{\boldmath$x$}}}=\frac{Ze^{-\xi_{\mbox{\boldmath$x$}}}}
{1+Ze^{-\xi_{\mbox{\boldmath$x$}}}}\\
\Pi_{\mbox{\boldmath$x$}}^i=-\frac{\partial}{\partial\zeta_i}\log\Xi_{\mbox{\boldmath$x$}}
&=&-\frac{mN_{\mbox{\boldmath$x$}}\zeta_{\mbox{\boldmath$x$}}^i}
{\beta_{\mbox{\boldmath$x$}}}.
\end{eqnarray*}

The hydrodynamic variables are the mass-density $\rho=mNa^{-3}$,
the velocity field
$\mbox{\boldmath$u$}=-\mbox{\boldmath$\zeta$}/\beta$, the
temperature $\Theta=\left(k_{_B}\beta\right)^{-1}$, and
the potential energy per unit mass, $\phi=\Phi/m$. We shall
therefore eliminate $\xi,\;\beta,\;\zeta^j$ in favour of
$\rho,\;\Theta,\;u^j$. Only $\xi$ remains to be found:
\begin{equation}
e^{-\xi_{\mbox{\boldmath$x$}}}=Z_{\mbox{\boldmath$x$}}^{-1}
N_{\mbox{\boldmath$x$}}/\left(1-N_{\mbox{\boldmath$x$}}\right),
\end{equation}
or
\begin{equation}
\xi=-\beta\Phi-\log N+\log(1-N)+\log Z_0
\end{equation}
where $Z_0$ is the canonical partition function when $\Phi=0$. We
need its gradient:
\begin{equation}
\partial_j\xi=\frac{3\partial_j\Theta}{2\Theta}-\frac{m}{2k_{_B}}
\frac{u^iu^i}{\Theta^2}\partial_j\Theta+\frac{mu^i}{k_{_B}\Theta}
\partial_ju^i-\frac{\partial_jN}{N}-\frac{\partial_jN}{1-N}-
\frac{\partial_j\Phi}{k_{_B}\Theta}+\frac{\Phi}{k_{_B}}
\frac{\partial_j\Theta}{\Theta^2}. \label{gradxi}
\end{equation}
We denote by $E,N$ and {\boldmath$\Pi$} the total values of the
mean energy, number and momentum; then (\ref{entropy}) gives for
the entropy at equilibrium
\begin{equation}
\Theta S(\mu)=E+k_{_B}\Theta\xi N-\mbox{\boldmath$u\cdot\Pi$}+
k_{_B}\Theta\log\Xi.
\end{equation}
Compare this with the thermostatic formula
\begin{equation}
\Theta S=E+k_{_B}\Theta\xi N-\mbox{\boldmath$u\cdot\Pi$}+PV
\end{equation}
where $P$ is the pressure and $V$ is the volume; we see that
\begin{equation}
P=k_{_B}\frac{|\Lambda|}{V}\Theta\log\Xi=-k_{_B}\Theta
a^{-3}\log(1-N).
\label{hcpressure}
\end{equation}

\section{The fundamental equation}
Our model for the dynamics is a path $\{\mu(t)\}_{t\geq 0}$
through $\Sigma(\Omega)$ determined by the Liouville motion,
interrupted by a thermalization at random points {\boldmath$x$},
occurring at random times. After a thermalization at
{\boldmath$x$}, the state changes to one in which observables at
{\boldmath$x$} are statistically independent of those at any
{\boldmath$y\neq x$}, and, restricted to
$\Omega_{\mbox{\boldmath$x$}}$, the state is in thermodynamic
equilibrium. The law for the random time between collisions is of
exponential form, but the rate of the process depends on the local
density of the gas, and its temperature, and so on the state
itself. This means that the dynamics falls outside the usual
theory of Markov chains, in which the updated state is linear in
the current state, and the transition matrix is the same for all
states; we are in the business of non-linear Markov chains.
Nonlinearity itself is not the problem; the ultimate goal of this
work is to put an external potential $\Phi$ into our version
\cite{RFS1} of the Navier-Stokes equations, which are non-linear.
However, we are going to use the concept of conditional
probability to derive the master equation, and care is needed if
we stray  from a path in $\Omega$ to a path in $\Sigma(\Omega)$
which does not come from a path in $\Omega$. We shall adopt the
following way out, which makes use of the assumption that the
density is low, and so the collision probability is small. In
calculating the probability per unit time that a configuration
$\omega$ at time $t\in(t_0-t,t_0)$ move to another point of
$\Omega$, we shall assume that the hydrodynamic parameters
satisfies the Euler equations, (\ref{eudensity})-(\ref{euenergy})
in the small time interval $(t_0-T,t_0)$. The initial values of
the hydrodynamic parameters in these equations are taken to be
those of the true state at time $t_0-T$. Here we only need to
consider $T<20 t_\ell$, since the survival probability falls
exponentially. We then assume that the Markov process between
$t_0-T$ and $t_0$ is linear as usual, and takes place in an
ambient Euler fluid which determines the rate of thermalization.
In this paper, we show that, neglecting collisions, the means of
the slow variables do indeed satisfy the Euler equations, showing
self-consistency. We can then show that there is no production of
entropy in this case.

The Liouville dynamics is determined as follows. If at time $t=0$
a configuration $\omega\in\Omega$ has a particle at
{\boldmath$x$}, then it follows Newton's laws
\begin{equation}\label{Newton}
\frac{d\mbox{\boldmath$x$}}{dt}=\frac{\mbox{\boldmath$k$}}{m},\hspace{.5in}
\frac{d\mbox{\boldmath$k$}}{dt}=-\nabla\Phi:=m\mbox{\boldmath$f$}
\end{equation}
for a time. Before we take the continuum limit, space is
$\Lambda$, which is discrete; a particle following (\ref{Newton})
will nearly always leave the lattice. In that case, after any
given random time $t$ after which a thermalization occurs, we
place the resulting thermalized particle at the lattice site
nearest to $\mbox{\boldmath$x$}(t)$, ties being decided by tossing
a coin. In the limit $a\rightarrow 0$ we expect this to introduce
a negligible error in the location being assigned to the particle.

Let $\mu\in\Sigma(\Omega)$ and define
$N_{\mbox{\boldmath$x$}}=\mu\{\omega_{\mbox{\boldmath$x$}}\neq\emptyset\}$,
the probability that {\boldmath$x$} is occupied in the state
$\mu$. Then $N_{\mbox{\boldmath$x$}}=\mu\cdot{\cal
N}_{\mbox{\boldmath$x$}}$. Let
$p_{\mbox{\boldmath$x$}}(\mbox{\boldmath$k$})$ be the conditional
probability that
$\omega_{\mbox{\boldmath$x$}}=\mbox{\boldmath$k$}$, given that
${\cal N}_{\mbox{\boldmath$x$}}=1$:
\begin{equation}
p_{\mbox{\boldmath$x$}}(\mbox{\boldmath$k$})=\mu
\{\omega:\omega_{\mbox{\boldmath$x$}}=
\mbox{\boldmath$k$}|\omega_{\mbox{\boldmath$x$}}\neq\emptyset\}=
N_{\mbox{\boldmath$x$}}^{-1}\mu\{\omega:\omega_{\mbox{\boldmath$x$}}=
\mbox{\boldmath$k$}\}.
\end{equation}
Thus,
\begin{equation}
\mu\{\omega:\omega_{\mbox{\boldmath$x$}}=\mbox{\boldmath$k$}\}=
N_{\mbox{\boldmath$x$}}p(\mbox{\boldmath$x,k$}).
\end{equation}
This does not mean of course that ${\cal N}_{\mbox{\boldmath$x$}}$
and {\boldmath${\cal P}_x$} are independent in the state $\mu$.

Given that no collisions occur, the particles obey (\ref{Newton}),
and this induces the Liouville motion on the states, namely,
Boltzmann's equation with no collision term:
\begin{equation}
\frac{\partial \mu(\mbox{\boldmath$x,k$})}{\partial
t}+\frac{\mbox{\boldmath$k$}}{m}\cdot\mbox{\boldmath$\partial$}\mu(\mbox
{\boldmath$x,k$})+\mbox{\boldmath$f\cdot\nabla_k$}\mu(\mbox{\boldmath$x,k$},t)
=0.
\end{equation}
This dynamics of the probabilities does not correspond to an
underlying motion in $\Omega$ for all time. We can find initial
conditions for which two particles both arrive within a distance
$a/2$ from the same point at the same time. Such a configuration
does not lie in $\Omega$. Our assumptions of no collisions is true
with high probability for a few time steps, but collisions are
almost sure to occur eventually. We will replace the problem of
tracking which collisions actually occur if we follow
(\ref{Newton}) by Boltzmann's idea of introducing a probability
for collisions. Instead of giving, as Boltzmann did, the
probability that a pair of particles with given position and
momentum produce another specified pair, it will be enough to
assume that the collision is 100\% efficient (in the terminology
of \cite{Rivet}). This means that after a collision at
{\boldmath$x$}, of a particle with momentum {\boldmath$k$}, we
replace this configuration by a particle at {\boldmath$x$} with
momentum $\mbox{\boldmath$k$}^\prime$, with probability determined
by the Maxwell distribution of mean momentum {\boldmath$k$} and
mean energy $\mbox{\boldmath$k\cdot k$}/(2m)+\Phi$. By
construction, this does not alter the means of the slow variables.
We shall refer to this event as a thermalization, rather than as a
collision. We shall assume that the mean free path is large on the
molecular scale, and neglect the possibility that a snapshot of
the lattice catch a particle in the process of thermalizing: with
high probability it will be in Newtonian motion between
collisions. From any initial configuration $\omega$, we can follow
the dynamics as time progresses, following this dynamics. It is a
random path through $\Omega$, that is, a process, defined for
$t_0-T\leq s\leq t_0$, and described by the family of points
$\{\mu(s)\}_{t_0-T\leq s\leq t_0}$.

Let $t$ denote the random time between collisions, whose law is
determined by the conditional probability
$w(\mbox{\boldmath$x,k$},t_0;t)dt$ that a particle, certainly at
{\boldmath$x$} at time $t_0$ with momentum {\boldmath$k$}, travel
under Newton's laws a free time $t$ and thermalize in the interval
$(t_0+t,t_0+t+dt)$. Since it will thermalize eventually,
\begin{equation}
\int_0^\infty w(\mbox{\boldmath$x,k$},t_0;t)dt=1.\label{normalize}
\end{equation}
The important small parameter is the mean free time, also called
the relaxation time, $t_\ell$:
\begin{equation}
t_\ell(\mbox{\boldmath$x,k$},t_0):=\int_0^\infty
t\,w(\mbox{\boldmath$x,k$},t_0;t)\,dt.\label{freetime}
\end{equation}
This is invariant under ${\cal G}$, unlike the mean free path,
which is related by $\ell=\mbox{\boldmath$k$}t_\ell/m$. For free
particles, $\Phi=0$, these were found in \cite{RFS1}. They are not
affected by the potential, so we use the same values here. We also
introduced
\begin{eqnarray}
C(\mbox{\boldmath$y,k$},t)dt&=&{\rm Prob}\{\mbox{a particle
at ({\boldmath$y,k$}) at time }t\mbox{ thermalize in }(t,t+dt)\}\\
W(\mbox{\boldmath$x,k$},t_0;t)&=&{\rm Prob}\left\{\mbox{a particle
at }(\mbox{\boldmath$x,k$},t_0)\mbox{ survive up to time }t_0
+t\right\}.
\end{eqnarray}
Put $z(t):=(\mbox{\boldmath$x$}(t),
\mbox{\boldmath$k$}(t)):=\tau_t(z)$, the solution to Newton's
equations with initial point $z:=(\mbox{\boldmath$x,k$})$. Then we
have the relation
\begin{equation}
w(\mbox{\boldmath$x,k$},t_0;t)=W\left(\mbox{\boldmath$x,k$},t_0;t\right)
\,C\left(z(t),t+t_0\right)
\end{equation}
and just as in \cite{RFS1} we get from (\ref{normalize}),
\begin{equation}
W(\mbox{\boldmath$x,k$},t_0;t)=\exp\left\{-\int_0^t
C(z(t_1),t_0+t_1)\,dt_1\right\}.
\end{equation}
As in \cite{RFS1}, eq.~(36), $C$ is proportional to the density
$\rho$. Thus if $\rho$ is bounded below by a positive number, we
get an exponential decrease for $W$, the survival probability,
along an orbit.

The fundamental equation will relate the probability distribution
$\mu(z,t_0)$ at an arbitrary point $z$ in phase space at the time
$t_0$ to a Maxwell distribution $\overline{\mu}$ at the same
point, with small correction terms. Let $z(s)$ denote (for $s>0$)
the point in phase space on the backward Newtonian orbit of the
point $z$ at time $t_0-s$. For each sample path
$\omega(\mbox{\boldmath$\cdot$})$, in which $z$ is occupied at
time $t_0$, there is a unique time $t_0-t$ at which the last
thermalization occurred, in that no further thermalizations took
place on the orbit $\{z(s)\}_{0<s<t}$. It follows that the free
time of the particle thermalized at $t_0-t$ must have a free time
$t^\prime$ say, with $t^\prime>t$.

In the following, `event' will mean an event in the sample space
(the path space of $\Omega$) of the process constructed above in
the time interval $(t_0-T,t_0)$. Let $E(s)$ denote the event, that
there is a particle at $z(s)$ at time $t_0-s$. We shall consider
the conditional probabilities of collision and free motion along
the orbit, given $E(0)$. We shall then recover a formula for
$\mu(z(0))$ using Bayes's theorem. Liouville's theorem, that
$d^3x\,d^3k$ is invariant under Newton's laws, enables us to
deduce equations relating the density of probability from
equations relating probabilities. For example we shall write
$\Pr\{E(0)\}=\mu(\mbox{\boldmath$x,k$},t_0)$, referring to the
densities.

Let $F(t)$ be the event, that $E(0)$ occurred, and the last
collision occurred at time $t_0-t$ producing a particle at the
phase point $z(t)$. Then $F(t)\subseteq E(0)$, since the event
produced exactly the right state, $z(t)$, which evolves to our
point $z(0)$, as it undergoes no further collisions. Let $F(a,b)$
denote the event, that the last collision was at some $s$, $a\leq
s\leq b$. If $(a,b)\cap(c,d)=\emptyset$, then $F(a,b)$ and
$F(c,d)$ are disjoint. So, assuming smoothness, $F(t)$ has a
probability density on ${\bf R}^+$, say $f(t)$. Let
$G(t,t^\prime)$ be the event that $z(t)$ is occupied at time
$t_0-t$, and that the particle's free time is $t^\prime$. Let
$H(t_1)$ be the event that $F(s)$ occurred for some $s<t_1$. We
shall need the formula
\begin{eqnarray}
\Pr\{F(t)\}/\Pr\{H(t^\prime)\}&=&f(t)/\int_0^{t^\prime} f(s)ds\nonumber\\
&=&\frac{1}{t^\prime}\left(1+\frac{f^\prime(0)}{f(0)}(t-t^\prime/2)\right)
+O(t^\prime),\label{lemma}
\end{eqnarray}
which is true if $f$ is smooth and $f(0)$ is not zero. We see that
$G(t,t^\prime)\cap H(t^\prime)\subseteq E(0)$. Hence certainly
\[
G(t,t^\prime)\cap F(t)\cap H(t^\prime)\subseteq E(0).\] Hence
\[
\Pr\{G(t,t^\prime)\cap F(t)\cap H(t^\prime)\cap
E(0)\}=\Pr\{G(t,t^\prime) \cap F(t)\cap H(t^\prime)\}.\] Thus
\begin{eqnarray*}
\Pr\left\{G(t,t^\prime)\cap F(t)\cap H(t^\prime)\raisebox{-.2cm}
{\rule{.01cm}{.6cm}}\;E(0)\right\}&:=&
\Pr\{E(0)\}^{-1}\Pr\{G(t,t^\prime)\cap F(t)\cap H(t^\prime)\cap E(0)\}\\
&=&\mu(z(0),t_0)^{-1}\Pr\{G(t,t^\prime)\cap F(t)\cap
H(t^\prime)\}.
\end{eqnarray*}
Let $\Pr_{H^\prime}\{\mbox{\boldmath$.$}\}$ denote the conditional
probability of an event, given $H(t^\prime)$. Then from what we
have just seen,
\begin{eqnarray*}
\Pr_{H^\prime}\left\{G(t,t^\prime)\cap
F(t)\raisebox{-.2cm}{\rule{.01cm}
{.6cm}}\,E(0)\right\}&:=&\Pr\left\{G(t,t^\prime)\cap F(t)\cap
H(t^\prime)
\raisebox{-.2cm}{\rule{.01cm}{.6cm}}\,E(0)\right\}\Pr\{H(t^\prime)\}^{-1}\\
&=&\Pr\{G(t,t^\prime)\cap F(t)\cap
H(t^\prime)\}\mu^{-1}\Pr\{H(t^\prime)\} ^{-1}.
\end{eqnarray*}
Now,
\[\int_0^\infty dt^\prime\int_0^{t^\prime}dt \Pr_{H^\prime}\left\{F(t)\cap
G(t,t^\prime)\raisebox{-.2cm}{\rule{.01cm}{.6cm}}\,E(0)\right\}=1.
\]
as the last collision must have happened for some $t$ and some
$t^\prime>t$. Hence
\[
\int_0^\infty dt^\prime\int_0^{t^\prime} dt \Pr\{G(t,t^\prime)\cap
F(t)\cap H(t^\prime)\}\mu^{-1}\Pr\{H(t^\prime)\}^{-1}=1.\] As
$\mu(z(0),t_0)$ does not depend on $t$ or $t^\prime$, we get
\[
\mu(z(0),t_0)=\int_0^\infty
dt^\prime\int_0^{t^\prime}dt\frac{\Pr\{G(t,t^\prime)\cap F(t)\cap
H(t^\prime)\}}{\Pr\{H(t^\prime)\}}.\] We can omit $H$ from the
numerator, as $t<t^\prime$ is enforced by the region of
integration. So
\begin{eqnarray}
\mu&=&\int_0^\infty
dt^\prime\int_0^{t^\prime}dt\frac{\Pr\{G(t,t^\prime)
\cap F(t)}{\Pr\{H(t^\prime)\}}\nonumber\\
&=&\int_0^\infty dt^\prime\int_0^{t^\prime}dt
\frac{\Pr\{G(t,t^\prime)
|\,F(t)\}\Pr\{F(t)\}}{\Pr\{H(t^\prime)\}}. \label{mu}
\end{eqnarray}
Also, if $t<t^\prime$, then
\begin{equation}
\Pr\left\{G(t,t^\prime)\raisebox{-.2cm}{\rule{.01cm}{.6cm}}\,F(t)\right\}
=\overline{N}\overline{p}(z(t),t_0-t) w(z(t),t_0-t;t^\prime),
\label{mubar} \end{equation} (what we have been calling a
collision is a thermalization.) In applying (\ref{lemma}), we note
that the case $f(0)=0$ corresponds to no collisions, so we may
assume that $f(0)\neq 0$. We expand (\ref{mubar}) up to first
order in $t$, as $t\leq t^\prime$, and $t^\prime\leq T=20
t_\ell=O(t_\ell)$; this gives
\[\Pr\left\{G(t,t^\prime)\raisebox{-.2cm}{\rule{.01cm}{.6cm}}
\,F(t)\right\}=\overline{\mu}(z,t_0)w(z,t_0;t^\prime)-t
\left[\partial_ik_i/m
+\partial_0-(\partial_j\Phi)\frac{\partial}{\partial k_j}\right]
\overline{\mu}(z,t_0)w(z,t_0;t^\prime).\] Put this, and use
(\ref{lemma}), in (\ref{mu}) to get
\begin{eqnarray}
\label{fund1} \mu(z,t_0)&=&\int_0^\infty
dt^\prime\left[\rule{0cm}{.6cm}\overline{\mu}(z,t_0)
w(z,t_0,t^\prime)\right.-\nonumber\\
&-&\left.t^\prime/2\left(\frac{k_i\partial_i}{m}+\partial_0-
(\partial_j\Phi)\frac{\partial}{\partial k_j}\right)\overline{\mu}
(z,t_0)w(z,t_0;t^\prime)\rule{0cm}{.6cm}\right].
\end{eqnarray}
The unknown term involving $f^\prime/f$ does not appear, because
$\int_0^{t^\prime}(t-t^\prime/2)dt$ vanishes, and
$\int_0^{t^\prime}(t-t^\prime/2)t\,dt$ is of second order in
$t^\prime$, and so can be neglected. We can now do the integral
over $t^\prime$, using (\ref{normalize}), and (\ref{freetime}).
This gives
\begin{equation}\label{fund2}
\mu=\overline{\mu}-\frac{1}{2}\left(\frac{k_j\partial_j}{m}+\partial_0
-\left(\partial_j\Phi\right)\frac{\partial}{\partial k_j}\right)
\left(\overline{\mu}\,t_\ell\right)
\end{equation}
This equation was derived in \cite{RFS1} using another method (aka
guesswork), in the case when $\Phi=0$.

It does not seem fruitful to seek accuracy up to and including
$O(t_\ell^2)$. This would involve introducing unknown parameters,
such as $f^\prime/f$; worse, we would have to keep terms of second
order in taking the continuum limit of the lattice; this
introduces diffusion terms into the equations, which come from the
Ito corrections to calculus. A similar extension of the work of
Chapman and Cowling \cite{Chapman}, who start with the Boltzmann
equation rather than a master equation, is generally agreed not to
have been worth the effort. Keeping only terms of first order in
$t_\ell$ leads to the surprising result that the equations of
motion are determined, knowing only that $\overline{\mu}$ is in
{\em LTE}. We do not need to assume, as is done in \cite{Balescu},
that the means of the slow variables in the approximating {\em
LTE}-state $\overline{\mu}$ are the same as the true means, in the
state $\mu$. Indeed, this turns out not to be the case; the reason
is that $\overline{\mu}$ is conditioned by the fact that a
thermalisation has occurred, and is not just the maximum entropy
estimate of $\mu$. Then $\mu(\mbox{\boldmath$x$},t_0)$ is the sum
of such terms from various nearby points
$(\mbox{\boldmath$x-k$}t/m,t_0-t)$ and for a given {\boldmath$k$},
all the contributions are from one side of {\boldmath$x$}, so the
means should differ unless the state is constant in space and
time. In this paper we neglect the difference between $\mu$ and
$\overline{\mu}$, which is the cause of dissipation in the
Navier-Stokes equations. We arrive at the Euler equations, and
show that they are free of dissipation, as expected.
\section{The Euler equations}
The Euler equations follow from the approximation of zero$^{\rm
th}$ order, in which the difference between $\mu$ and
$\overline{\mu}$ is neglected. Mean fields for states in {\em LTE}
are computable in terms of Gaussian integrals.

The velocity field of a particle is the random field
$\Upsilon_{\mbox{\boldmath$x$}}:=\mbox{\boldmath${\cal
P}$}_{\mbox{\boldmath$x$}}/m$, and the mean current of a random
variable $\chi$, conserved or not, is, in the continuum,
low-density limit
\begin{equation}
{\bf J}_\chi=\int
d^3k\,\mu\mbox{\boldmath$\Upsilon$}\chi.\label{flow}
\end{equation}
Our assumption is that on thermalization there is no change in the
means of ${\cal N}_{\mbox{\boldmath$x$}}$, {\boldmath${\cal P}_x$}
or ${\cal E}_{\mbox{\boldmath$x$}}$. Since the space integrals of
$N_{\mbox{\boldmath$x$}}$ and $E_{\mbox{\boldmath$x$}}$  are to be
conserved in time, their equations of motion are of the form
\begin{eqnarray}
\dot{\rho}&+&\partial_j J_\rho^j=0\label{conservmass}\\
\dot{E}&+&\partial_jJ_{_E}^j=0.\label{conservenergy}
\end{eqnarray}
Momentum is not conserved; in a small volume at {\boldmath$x$} a
particle of momentum {\boldmath$k$} in time $dt$ is changed to
$\mbox{\boldmath$k$}-\mbox{\boldmath$\nabla$}\Phi\,dt$, so the
momentum obeys
\begin{equation}
\frac{d\varpi^i}{dt}+\partial_jJ_{\varpi^i}^j+\frac{\rho}{m}
\partial_j\Phi=0.
\label{nonconservmom}
\end{equation}
In \cite{RFS1} we showed that we can evaluate the currents
(\ref{flow}) of hydrodynamics exactly if $\mu$ is in {\em LTE},
with parameters $\rho, u^i,\Theta$ say. By the same method,
(\ref{conservmass})-(\ref{nonconservmom}) can be written out:
\begin{eqnarray}
\frac{d\rho}{dt}&+&\partial_j\left(\rho u_j\right)=0\label{eudensity}\\
\frac{d}{dt}\left(\rho u_i\right)&+&\partial_j\left(\rho
u_iu_j\right)+\partial_iP+
\rho\partial_i\phi=0\label{eumomentum}\\
\frac{d}{dt}\left\{\rho\left(\frac{3k_{_B}}{2m}\Theta+\frac{1}{2}u_ju_j+
\frac{\Phi}{m}\right)\right\}&+&\partial_j\left\{\rho
u_j\left(\frac{3k_{_B}}{2m}\Theta+P/\rho+\frac{1}{2}u_ju_j+\phi\right)\right\}=0.
\label{euenergy}
\end{eqnarray}
Here, $\phi=\Phi/m$. Note that the pressure $P$ appearing in these
equations is that of an ideal gas, not the equilibrium pressure of
the model; the two differ by terms of second order in the density.
This reflects that the derivation of the fundamental equation
depends on the assumption that the density is small.
\section{Entropy production}
From von Neumann's formula,
\begin{equation}
S:=-\sum_\omega \mu(\omega)\log\mu(\omega)
\end{equation}
we conclude that
\begin{equation}
\dot{S}=-\sum_\omega
\dot{\mu}(\omega)\log\mu(\omega)-\sum_\omega\dot{\mu}(\omega).
\end{equation}
The last term is zero, so we may regard $-\log\mu$ as a random
variable ${\cal S}$ whose average rate of change is given in the
Schr\"{o}dinger picture by $\dot{\mu}\cdot{\cal S}$. The current
of the entropy would then be, per site
\begin{equation}
a^3{\cal J}_{_S}^j(x)={\cal S}(\omega)\Upsilon^j=\xi{\cal
N}\Upsilon^j+\beta{\cal E}\Upsilon^j+\zeta^i{\cal
P}_i\Upsilon^j+\log\Xi\;\Upsilon^j,
\end{equation}
when $\mu=Np$ and $p$ is in {\em LTE}. Its mean density, $J_{_S}^j$, in
the state $\mu$ has three terms:
\begin{eqnarray*}
\xi a^{-3}\mu\cdot\left({\cal N}\Upsilon^j\right)+a^{-3}\log\Xi\mu\cdot
\Upsilon&=&\frac{\xi}{m}J_\rho^j-\log(1-N)u^j\\
a^{-3}\zeta^i\mu\cdot\left({\cal P}_i\Upsilon^j\right)&=&\zeta^iJ_{\varpi_i}
^j\\
a^{-3}\beta\mu\cdot\left({\cal E}\Upsilon^j\right)&=&\beta J_{_E}^j.
\end{eqnarray*}
We expect that entropy will be conserved, because of Kossakowski's
argument \cite{Kossakowski}: in the absence of collisions, the
projections onto the information manifold do not create any
entropy if time is continuous. See also \cite{Balian}. Without
collisions, Boltzmann's equation also give zero entropy
production, if the assumption of {\em LTE} is made. So we expect
that the rate of change of entropy is balanced by the outflow
though the boundary:
\begin{equation}
\dot{s}+\partial_j\left(\frac{\xi}{m}J_\rho^j+\zeta^iJ_{\varpi_i}^j+\beta
J_{_E}^j-\log(1-N)u^j\right)=0. \label{conservation}
\end{equation}
where $s$ denotes the entropy density. On the other hand, we have
\[
a^3s=\xi N+\zeta^i\Pi_i+\beta E+\log\Xi,\] giving
\begin{eqnarray}
a^3\frac{ds}{dt}&=&N\partial_t\xi +\Pi_i\partial_t\zeta^i+
E\partial_t\beta+\frac{1}{\Xi}\left(
\frac{\partial\Xi}{\partial\xi}\partial_t\xi
+\frac{\partial\Xi}{\partial\zeta^i}\partial_t\zeta^i
+\frac{\partial\Xi}{\partial\beta}\partial_t\beta\right)\label{firstline}\\
& & +\xi \partial_t N+\zeta^i\partial_t\Pi_i+\beta\partial_t
E\label{secondline}.
\end{eqnarray}
Now,
\[
\frac{1}{\Xi}\frac{\partial \Xi}{\partial\xi}=-N \mbox{ etc.}\] so
the first line (\ref{firstline}) vanishes, to leave
\[
a^3\partial_t s=\xi\partial_t
N+\zeta^i\partial_t\Pi_i+\beta\partial_t E.\] We divide by $a^3$
to get the densities $\rho$, $a^{-3}E$ and $\varpi_i$, and the
relation
\begin{eqnarray}
\int_\Lambda \partial_t s\,d^3x&=&\int_\Lambda\left(
\frac{\xi}{m}\partial_t\rho+\zeta^i\partial_t\varpi_i+\beta\partial_t
E/a^3\right)d^3x\\
&=&-\int_\Lambda\left( \frac{\xi}{m}\partial_jJ_\rho^j
+\zeta^i\left(\partial_jJ_{\varpi_i}^j+\rho\delta_{ij}\partial_j
\phi\right)+\beta\partial_jJ_{_E}^j\right)\nonumber\\
&=&\int_\Lambda \left(\zeta^i\rho f_i+\partial_j\xi
J_\rho^j+\partial_j\zeta^iJ_{\varpi_i}^j+\partial_j\beta J_{_E}^j\right)
\label{ons1}\\
&-&\oint_{\partial\Lambda}\left(\xi
J_\rho^n+\zeta^i(J_{\varpi_i}^n+\delta_{ni}\Phi/m)+\beta
J_{_E}^n\right)d\sigma_n
\label{ons2}
\end{eqnarray}
where $n$ denotes the normal direction to the boundary. The term
in (\ref{ons2}) involving $f_i=-\partial_i\Phi/m$ represents the
free-energy change due to the work done by the external field on the
particles; we shall see that this is cancelled out by another term,
showing that this work done is not thermalised by the dynamics. In spite of
the apparent Onsager form of (\ref{ons1}),
we cannot identify it as the entropy production and (\ref{ons2})
as the flow through the boundary, because the surface term differs
from the entropy current,
as in (\ref{conservation}) by $-\log(1-N)u^j$. Indeed, if we put
the Euler currents in (\ref{ons1}) we do not get zero, even when
$\Phi=0$. We must therefore add and subtract the missing term, to
get
\begin{eqnarray}
\partial_t s&=&\int_\Lambda\left(\zeta^i\rho f_i+\partial_j\xi
J_\rho^j+\partial_j\zeta^iJ_{\varpi_i}^j+\partial_j\beta
J_{_E}^j-\partial_j(a^{-3}\log(1-N)u_j)\right)d^3x\label{ons3}\\
&&-\oint_{\partial\Lambda}J_{_S}^id\sigma_i.
\end{eqnarray}
If we now put in the Euler currents,
(\ref{eudensity})-(\ref{euenergy}), we still do not find exactly
zero for the entropy production, (\ref{ons3}). Indeed, the entropy
production involves the logarithm, whereas the Euler equations are
polynomial, so cancellation is not possible. We find that
(\ref{ons1})  vanishes up to terms $O(N^2)$, that is, in the limit
of low density. This reflects the low-density assumption inherent
in the axiom that the {\em BBGKY} currents, (\ref{flow}),  are the
actual currents of particles flowing out of the region $d^3x$.
While this is a reasonable model for point-like particles, it
neglects the fact that if the lattice-site just beyond the
boundary is occupied, then a particle moving out of the region
will not be able to land, as no configuration in the sample space
can have two particles at the same site. Our dynamics did not need
to specify the rule as to what will happen, as (in the low density
limit) the probability that the point is occupied is small.

It turns out that if we modify the pressure in the Euler
equations, to be the equilibrium pressure of the gas with hard
core, (\ref{hcpressure}), rather than that of the ideal gas, then
the entropy production is exactly zero, in conformity with
Kossakowski's theorem:

\begin{theorem}
{\em Take the pressure $P$ in (\ref{eumomentum}) and (\ref{euenergy}) to be
that of the hard-core gas at equilibrium, (\ref{hcpressure}). Then the
entropy production, (\ref{ons3}), is zero.}
\end{theorem}
{\bf PROOF}. We first recall the relation between the canonical
variables and the hydrodynamic variables. We saw that
\[
\beta=\frac{1}{k_{_B}\Theta}\hspace{1.5in}\zeta^j=-\beta
u^j=-\frac{u^j}{k_{_B}\Theta}.\] so from (\ref{gradxi}) we get for
the entropy production (\ref{ons3}): \begin{eqnarray*}
 \dot{s}_{\rm prod}&=&\zeta^i\rho f_i\\
 &+&\frac{\rho
 u^j}{m}\left(\frac{3}{2}\frac{\partial_j\Theta}{\Theta}-\frac{m}{2k_{_B}}
\frac{u^iu^i}{\Theta^2}\partial_j\Theta+\frac{m}{k_{_B}}\frac{u^i}{\Theta}
\partial_ju^i
-\frac{\partial_jN}{N}-\frac{\partial_jN}{1-N}-\frac{m}{k_{_B}}
\frac{\partial_j\phi}{\Theta}+\frac{m}{k_{_B}}\frac{\phi}{\Theta^2}
\partial_j\Theta\right)\\
&+&\left(\rho
u_iu_j+\delta_{ij}P\right)\left(-\frac{\partial_ju^i}{k_{_B}\Theta}
+\frac{u^i}{k_{_B}}\frac{\partial_j\Theta}{\Theta^2}\right)\\
&+&\rho
u_j\left(\frac{3k_{_B}}{2m}\Theta+P+\frac{1}{2}u^iu^i+\phi\right)\frac{
-\partial_j\Theta}{k_{_B}\Theta^2}\\
&+&a^{-3}\frac{\partial_j
N}{1-N}u^j-a^{-3}\log(1-N)\partial_ju^j\\
&=&0.
\end{eqnarray*}

\vspace{.1in} {\bf Acknowledgements}. This work summarizes my
contributions to the conference on `Nonlocal elliptic and parabolic
 problems', Bedlewo, and to the meeting of the Polish
Physical Soc., Gdansk, 2003. I thank P. Biler, G. Karch, T.
Nadzieja, R. Alicki and W. A. Majewski for arranging the visits.

\end{document}